\RequirePackage{fix-cm}
\documentclass{svjour3}                     
\smartqed  

\usepackage{graphicx}
\usepackage{multirow}
\usepackage{paralist}
\usepackage{setspace}
\usepackage{hyperref}
\usepackage{xurl}
\usepackage[numbers]{natbib}

\usepackage{orcidlink} 
\usepackage{amsmath}

\usepackage{siunitx}

\usepackage[linesnumbered, ruled, vlined]{algorithm2e}
\SetAlFnt{\small}
\SetAlCapFnt{\small}
\SetAlCapNameFnt{\small}
\SetKwRepeat{Do}{do}{while}
\usepackage{booktabs}
\usepackage{xcolor}
\usepackage{subcaption}
\usepackage{tcolorbox}
\newcommand{\rev}[1]{\textcolor{black}{#1}}

\begin{document}

\title{Detecting Multiple Semantic Concerns in Tangled Code Commits
}


\author{
Beomsu Koh\orcidlink{0009-0004-2742-2200}
\and
Neil Walkinshaw\orcidlink{0000-0003-2134-6548}
\and
Donghwan Shin\orcidlink{0000-0002-0840-6449}
}


\institute{
Beomsu Koh \at
School of Computer Science, University of Sheffield, Sheffield, United Kingdom \\
\email{gobeumsu@gmail.com}
\and
Neil Walkinshaw \at
School of Computer Science, University of Sheffield, Sheffield, United Kingdom \\
\email{n.walkinshaw@sheffield.ac.uk}
\and
Donghwan Shin \at
School of Computer Science, University of Sheffield, Sheffield, United Kingdom \\
\email{d.shin@sheffield.ac.uk}
}

\date{Received: date / Accepted: date}

\maketitle
\begin{abstract}

Code commits in a version control system (e.g., Git) should be \textit{atomic}, i.e., focused on a single goal, such as adding a feature or fixing a bug.
In practice, however, developers often bundle multiple concerns into \textit{tangled} commits, obscuring intent and complicating maintenance.
Recent studies have used Conventional Commits Specification (CCS) and Language Models (LMs) to capture commit intent, demonstrating that Small Language Models (SLMs) can approach the performance of Large Language Models (LLMs) while maintaining efficiency and privacy \rev{within local infrastructure}.
However, they do not address tangled commits involving multiple concerns, leaving the feasibility of using LMs for multi-concern detection unresolved.

In this paper, we frame multi-concern detection in tangled commits as a multi-label classification problem and construct a controlled dataset of artificially tangled commits based on real-world data.
We then present an empirical study using SLMs to detect multiple semantic concerns in tangled commits, examining the effects of fine-tuning, concern count, commit-message inclusion, and header-preserving truncation under practical token-budget limits.
Our results show that a fine-tuned \rev{27B}-parameter SLM \rev{outperforms} a state-of-the-art LLM \rev{across all concern counts}.
In particular, including commit messages improves detection accuracy by up to \rev{31\%} (in terms of Hamming Loss) with negligible latency overhead, establishing them as important semantic cues.
\end{abstract}
\section{Introduction}\label{sec:intro}

Commits are the basic unit of change in version control systems (VCS), combining code modifications (i.e., \emph{diffs}) and textual descriptions (i.e., \emph{commit messages}) to convey developer intent.
Modern software engineering practice encourages \emph{atomic commits}—small, coherent changes associated with a single, well-defined intent—because they facilitate code review, support debugging, and improve repository mining for research and automation \citep{Herzig2011Untangling}. 
In this work, we define an atomic commit as \emph{syntactically coherent} (i.e., consistent at the line, method, and file levels) and \emph{semantically cohesive} (i.e., addressing exactly one semantic concern, such as a feature addition, a bug fix, or a refactoring).

Building on this principle, the \emph{Conventional Commits Specification} (CCS) is widely used in industry as a lightweight standard for expressing commit intent (e.g., \texttt{feat} for adding features, \texttt{fix} for fixing bugs, \texttt{refactor} for refactoring, and \texttt{docs} for improving documentation) and encouraging structured, atomic commits \citep{ConventionalCommits2023}.
Despite such guidelines, empirical studies show that commit atomicity is often violated in practice, as under time pressure or when tasks overlap, developers frequently combine multiple concerns into a single commit, resulting in tangled commits \citep{Herzig2013Impact,Barnett2015Helping,Partachi2020Flexeme}, obscuring developer intent and complicating maintenance. 

Meanwhile, automated untangling has been studied based on heuristic and structural cues, such as partitioning by file paths \citep{Herzig2011Untangling,Herzig2013Impact}, dependency relations \citep{Barnett2015Helping,Shen2021SmartCommit}, abstract syntax tree (AST) proximity \citep{Muylaert2018Untangling}, and pattern templates mined from prior commits \citep{Kirinuki2014Hey}. 
These approaches are simple and interpretable, and they are effective at capturing \emph{syntactic} proximity.
\rev{Their prediction target, however, is a division of the changed code rather than a statement of the intent behind it, so the groups they produce carry no concern label.}
Subsequent work introduced machine-learning classifiers that exploit structural features, while more recent deep-learning models embed code changes into graphs or vectors for automated partitioning \citep{Dias2015Untangling,Partachi2020Flexeme,Li2022UTANGO,Fan2024Detect,Xu2025Detecting}. Although these techniques achieve higher accuracy than purely heuristic rules, \rev{they retain that same target, and naming the concerns a commit addresses remains a separate question}.

On the other hand, with advances in language models (LMs), recent studies increasingly consider \emph{semantic} concern based on code diffs and commit messages, and more recent work has begun to explore small language models (SLMs) for their computational efficiency and privacy \citep{Zeng2025First,Li2024Understanding}.
This line of work ranges from binary detection (e.g., identifying defect-prone changes) \citep{Lin2023CCT5} to CCS-oriented intent inference, where messages encode explicit intent and diffs serve as supporting evidence. 
However, most studies focus on single-label or binary classification, implicitly assuming one concern per commit and overlooking the fact that tangled commits often interleave multiple concerns \citep{Herzig2013Impact,Barnett2015Helping}.


To address this gap, we formulate the problem as \textit{multi-label classification of semantic concerns}, aiming to detect multiple concerns within tangled commits. \rev{Concern labels are actionable at the commit level even without attributing individual lines: flagging a commit that carries several concerns lets the author split it before it enters the project history, and applying the same detector across a repository's history reveals how often tangling occurs and which components are repeatedly involved, a form of evolution monitoring for which commit-level classification is already used~\citep{Sarwar2020Multilabel}.}

We focus on SLMs rather than large language models (LLMs), considering computational efficiency and data privacy \rev{within local infrastructure}.
For example, unlike LLMs, SLMs can often be executed locally on modern laptops, enhancing data privacy by keeping sensitive code and commit messages on-device rather than transmitting them to centralized servers \citep{Lu2025Small}. 

By focusing on SLMs, we aim to answer the following research questions.
\begin{enumerate}[\bf RQ1]
    \item \textit{What is the impact of the number of semantic concerns in a commit on multiple concern detection accuracy?} RQ1 evaluates how concern count—our operational proxy for atomicity and semantic complexity—affects multi-label detection accuracy and identifies the range in which a fine-tuned SLM remains a viable alternative to an LLM baseline.
    \item \textit{How much does semantic information from commit messages contribute to detection accuracy?} While diffs contain the code changes, commit messages provide additional semantic context about developer intent and the rationale behind those changes. We systematically evaluate the impact of commit-message inclusion on detection accuracy by conducting controlled experiments with and without messages.
    \item \textit{How robust is detection accuracy under realistic token-budget constraints when diffs are partially truncated using a header-preserving policy?} RQ3 evaluates the robustness of multi-concern detection under token-budget constraints using this header-preserving truncation policy.
    \item \textit{How do the same factors examined in RQ\numrange{1}{3} affect inference efficiency for fine-tuned SLMs?} While RQ\numrange{1}{3} focus on accuracy, practical deployment also depends on whether an SLM can process commits quickly enough. By measuring how these task characteristics influence end-to-end latency, we characterize the computational costs associated with each factor and identify which constraints impose the greatest slowdown during inference.
\end{enumerate}

Our results show that the fine-tuned SLM \rev{outperforms} an LLM baseline \rev{across all tested concern counts}.
We also find that commit messages consistently improve performance.
Under our header-preserving truncation policy, reducing the available token budget has only a limited effect on SLM accuracy and inference latency within the tested range. 
This suggests that, in our setting, huge context windows may not be necessary when commits fit within modest budgets and informative diff prefixes are retained.

To summarize, our key contributions are as follows:
\begin{enumerate}[1.]
    \item To the best of our knowledge, we provide the first systematic empirical study of the performance and efficiency of SLMs in detecting multiple concerns within tangled commits, analysing the impact of concern count, commit-message inclusion, and token-budget-constrained diff truncation.
    \item We release our replication package, including a carefully crafted benchmark dataset of 1750 synthesised tangled commits derived from a CCS-labelled real-world commit corpus \citep{Zeng2025First}, fine-tuned models based on Qwen3\rev{.6}-\rev{27B} \citep{Yang2025Qwen3}, and supporting scripts, to facilitate future research on commit untangling and semantic analysis (see Section~\ref{sec:data-availability}).
    \item We empirically demonstrate the usefulness of SLMs for multi-concern detection in tangled commits and identify key factors, including SLM fine-tuning and commit-message inclusion. We also show that large context windows are not always necessary within the tested token budgets and under our header-preserving truncation policy. 
\end{enumerate}

\section{Background}
\label{sc:background}
This section introduces the essential concepts that the rest of the paper builds on.
\subsection{Tangled Commits}
\label{sc:bg:tc}

In version control systems, a commit represents the smallest unit of change, combining code diffs and a commit message describing developer intent. 
An \emph{atomic commit} is a change that addresses a single developer intent~\citep{Muylaert2018Untangling,Li2022UTANGO,Fan2024Detect}.
However, developers frequently produce \emph{tangled commits} that intermix multiple concerns, often under time pressure or task overlap, which undermines clarity and complicates downstream tasks \citep{Fan2024Detect,Partachi2020Flexeme}.
Studies report that over 90\% of tangled commits involve three or fewer concerns~\citep{Shen2021SmartCommit,Herzig2013Impact} and characterize tangling as a pragmatic trade-off under delivery pressure~\citep{Herzig2016Impact}.

Accordingly, existing research has attempted to untangle commits automatically by analysing the structural relations among code changes.
Early heuristic methods partitioned edits based on file paths and dependency graphs, achieving interpretability but often yielding false positives—commits that appear atomic yet contain multiple developer intents~\citep{Herzig2011Untangling,Herzig2013Impact,Kirinuki2014Hey,Barnett2015Helping}.
Subsequent studies introduced machine learning classifiers with hand-engineered structural features to enhance separation accuracy~\citep{Dias2015Untangling,Partachi2020Flexeme}, as well as graph-partitioning methods that integrate program dependency information~\citep{Shen2021SmartCommit}.
More recently, deep learning approaches have embedded code changes into vector or graph representations, enabling scalable and automated untangling~\citep{Li2022UTANGO,Fan2024Detect,Xu2025Detecting}.
Despite these advances, existing approaches still treat atomicity through a structural lens, effectively equating it with syntactic cohesion.
These models learn expressive associations between commit messages and code changes, yet they often fail to recover the explicit developer intent underlying each change.

However, this persistent limitation arises from ambiguity in the definition of atomicity itself.
The traditional notion of being \emph{single-purpose} captures structural unity but neglects the semantic dimension of \emph{why a change is made}.
Atomicity, therefore, cannot be characterised solely in terms of a structural single-purpose criterion.
Instead, atomicity should be defined to encompass both syntactic and semantic cohesion.
An \emph{atomic commit} is therefore a change that is both syntactically and semantically cohesive, aligning structural unity with one consistent developer intent.

The CCS provides a practical means to map developer intent to explicit \emph{semantic concerns}.
Recent studies have leveraged language models capable of understanding both code and natural language to perform CCS-based classification, effectively identifying atomic commits through their expressed intent. 
This line of research naturally extends toward multiple-concern detection, where a single commit may embody several distinct semantic concerns. 
By enabling classification at the concern type level, CCS facilitates a more explicit capture of semantic intent—bridging the gap between structural cohesion and the developer’s underlying rationale for the change.

\rev{Semantic concern classification and commit untangling therefore answer different questions. Untangling partitions the changed code into structurally coherent groups, whereas concern classification reports which semantic concerns a commit contains as a whole. The two tasks have different prediction targets and outputs, so a partition cannot be scored against commit-level concern labels; \emph{concern attribution}, the mapping from each detected concern to the code fragments that realize it, is the component that would connect them. This study addresses commit-level detection and leaves attribution to future work.}

\subsection{Conventional Commits Specification  (CCS)}
\label{sc:bg:ccs}

\begin{figure}
    \centering
    \includegraphics[width=0.6\linewidth]{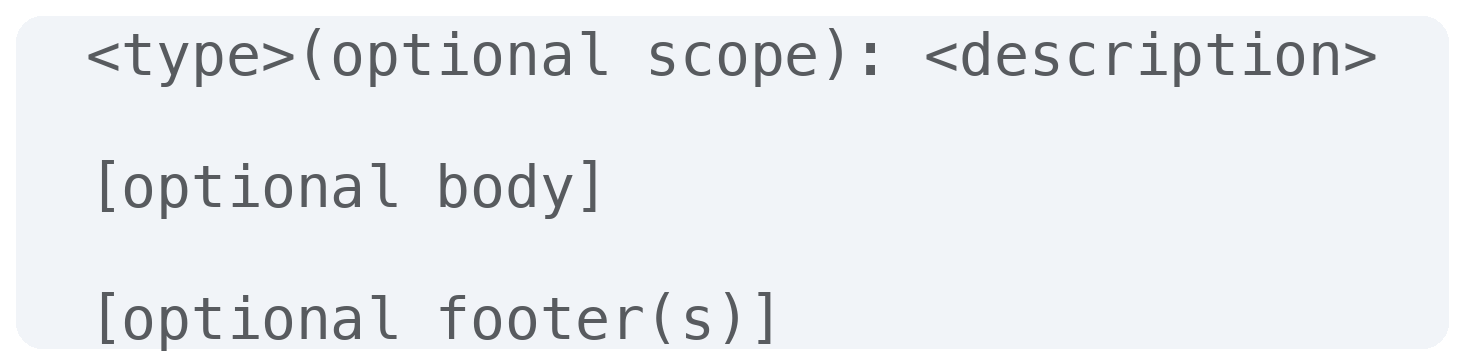}
    \caption{Conventional Commit Specification (CCS) format.}

    \label{fig:ccs-format}
\end{figure}

CCS defines the lightweight and uniform format for structuring commit messages~\citep{ConventionalCommits2023}.
As illustrated in Figure~\ref{fig:ccs-format}, a CCS message consists of a typed header followed by an optional body and footer. 
The header provides a concise description of the change and begins with a commit type indicating the developer’s intent.
It has recently become the most widely adopted commit convention in practice~\citep{Zeng2025First}.
The specification defines a set of commit types that describe the intent of code changes—such as \texttt{build}, \texttt{ci}, \texttt{docs}, \texttt{feat}, \texttt{fix}, \texttt{perf}, \texttt{refactor}, \texttt{style}, \texttt{test}, and \texttt{chore}.

\begin{figure}
    \centering
    \includegraphics[width=0.75\linewidth]{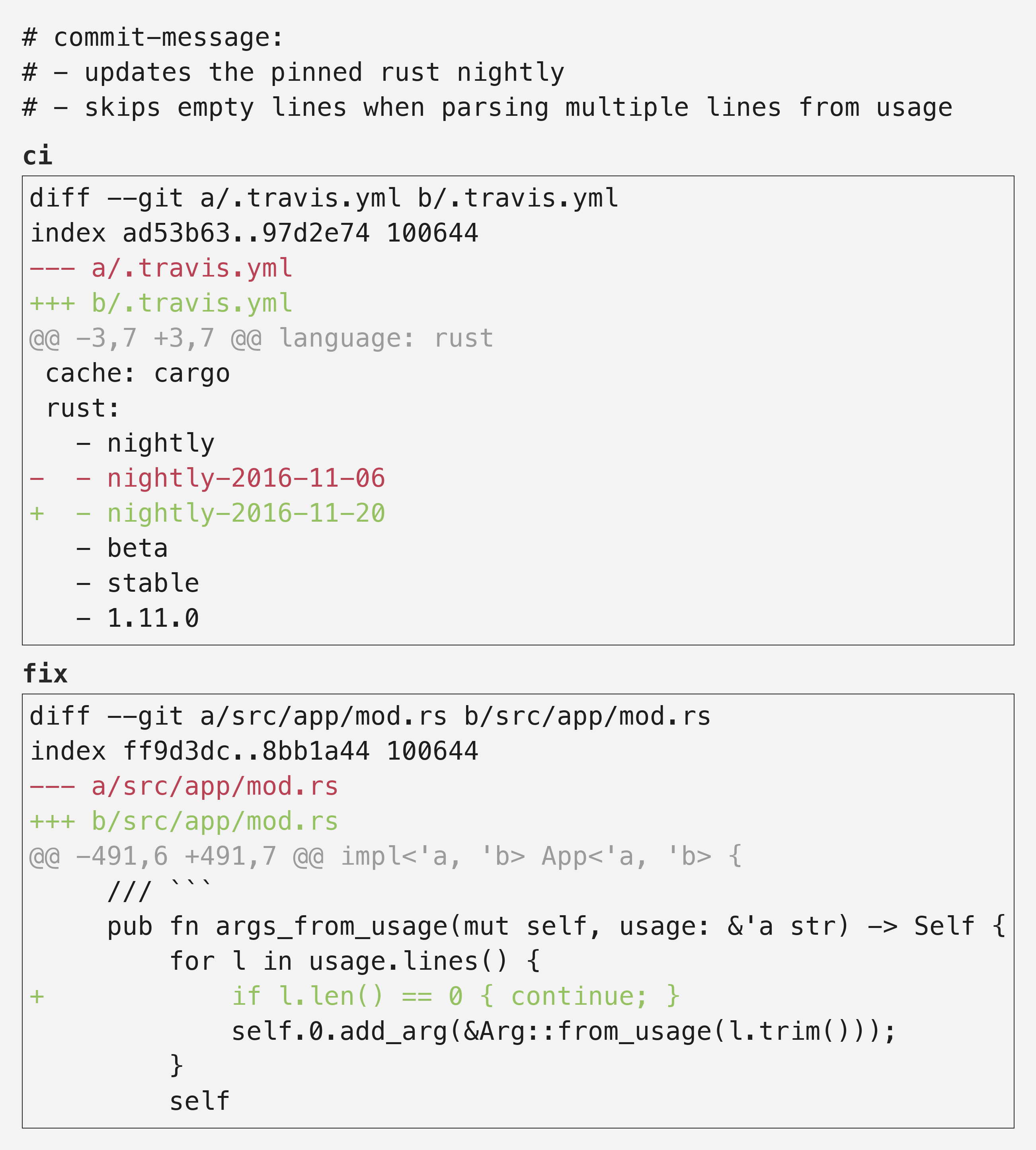}
    \caption{Motivating example illustrating \rev{a tangled commit with multiple} CCS types}
    \label{fig:motivating-example}
\end{figure}

In practice, these intent categories overlap in real commit histories.
\rev{Figure~\ref{fig:motivating-example} shows a tangled commit synthesised for our dataset as described in Section~\ref{sc:met:dc}.}
\rev{It contains both \texttt{ci} and \texttt{fix}.}
\rev{A developer might make both changes together for the practical reasons discussed in Section~\ref{sc:bg:tc} and commit them as a single unit.}
\rev{However, CCS records a single type per commit, so the commit is labelled either \texttt{ci} or \texttt{fix}, and either choice leaves the other concern unrecorded.}
This ambiguity motivates the need for clear classification guidelines.

Furthermore, prior studies report notable category overlap and semantic ambiguity within this taxonomy~\citep{Zeng2025First,Li2024Understanding}.
\citet{Zeng2025First} observed that CCS blends purpose-oriented types (\texttt{feat}, \texttt{fix}, \texttt{refactor}, \texttt{perf}, \texttt{style}) with object-oriented ones (\texttt{docs}, \texttt{test}, \texttt{ci}, \texttt{build}), which can lead to ambiguous cases.
They proposed prioritizing the purpose dimension when both dimensions coexist.
Meanwhile, \citet{Li2024Understanding} addressed this ambiguity by refining the taxonomy itself, removing \texttt{perf} and \texttt{chore}—labels that human annotators struggle to distinguish reliably—to improve consistency.

\subsection{Language Models}
\label{sec:background-lm}
Transformer architectures introduced the attention mechanism that underpins modern foundation models \citep{Vaswani2017Attention}.
Large language models (LLMs) such as GPT and LLaMA have shown strong effectiveness across software engineering tasks, including code generation and program repair~\citep{Fan2023Large,Hou2024Large}.
Despite their capabilities, LLMs impose substantial computational cost and latency, making them impractical for on-device or frequent deployment~\citep{Lu2025Small,Belcak2025Small}.
These cost, latency, and infrastructure demands have prompted growing interest in small language models (SLMs), which aim to retain practical utility while operating under modest resource budgets~\citep{Abdin2024Phi4,Yang2025Qwen3}.

Small language models (SLMs) are designed for efficient deployment on the modest hardware budgets.
However, the literature offers no single parameter threshold: some studies treat models around 5B parameters as SLMs~\citep{Lu2025Small}, others adopt 10B~\citep{Belcak2025Small}, and recent budget-trained variants extend to roughly 14B parameters~\citep{Abdin2024Phi4,Yang2025Qwen3}.
To reconcile these differences, \citet{Belcak2025Small} argue against a fixed parameter cutoff, defining SLMs instead by deployment feasibility and design intent.
They articulate three guiding principles:
\begin{itemize}[-]
\item \textbf{Timelessness}: Not limited by the hardware of a specific time.
\item \textbf{Practicality}: Designed to run on widely available consumer devices with fast inference.
\item \textbf{Motivation alignment}: Built to work well under limited resource budgets.
\end{itemize}

\rev{In line with this deployment-centred definition, the practical advantage of SLMs in our setting lies in local deployment and the resulting control over source-code data rather than in model size itself.}

The performance of SLMs is strongly influenced by the quality of their training data~\citep{Hou2024Large}.
Whereas LLMs can tolerate noisy, large-scale corpora, SLMs tend to learn more effectively from smaller and carefully curated datasets.
Prior studies~\citep{Abdin2024Phi4,Li2024Understanding} show that SLMs trained on expert-validated, high-quality datasets can outperform much larger models, underscoring that data curation is a primary determinant of performance.
Consequently, systematic filtering and expert validation are critical for building high-performing SLMs. 

Beyond data quality, recent advancements focus on enhancing the reasoning capabilities of SLMs to handle more complex tasks.
Chain-of-thought (CoT) prompting guides models to reason step by step, yielding better performance on tasks such as code generation and debugging, without modifying the model architecture~\citep{Fan2023Large,Hou2024Large,Abdin2024Phi4}.
While early studies suggested that such reasoning capabilities were emergent abilities exclusive to large models, recent work demonstrates that they can be effectively transferred to SLMs through knowledge distillation~\citep{Ranaldi2024Aligning}.
By aligning with the reasoning processes of larger teacher models, SLMs can learn to perform multi-step reasoning and code generation tasks with high accuracy.

Fine-tuning is also essential for domain specialization of SLMs, but full retraining is computationally impractical. 
Parameter-efficient fine-tuning (PEFT)~\citep{Liu2023Empirical}, which updates only a small subset of parameters while keeping the base model largely frozen, overcomes this limitation.
Among PEFT methods, low-rank adaptation (LoRA)~\citep{Hu2021LoRA} provides efficient task adaptation and rapid task switching without increasing inference latency or degrading model quality~\citep{Liu2023Empirical}.
By reducing computational and memory overhead, LoRA enables fine-tuning on consumer-grade hardware and can match or surpass larger models in software-engineering tasks; for example, a LoRA-tuned CodeLlama-7B achieved a 7\% accuracy gain over GPT-4 in a recent evaluation.
\section{Methodology}
\label{sc:methodology}

This section outlines the methodology used in this study.
We first refine the CCS labels to remove ambiguity and then construct a tangled commit dataset for evaluation.
In our experiments, each model receives a tangled commit as input and predicts a set of CCS labels representing its semantic concerns.

\subsection{CCS Refinement}
\label{sc:met:ccs}
We use CCS types because the CCS taxonomy provides an explicit mapping from a developer’s underlying intent to the semantic concerns reflected in the resulting code changes.
While the intent describes why a modification was made, the corresponding concern captures how that intent materializes within the codebase.

As described in Section~\ref{sc:bg:ccs}, the original CCS taxonomy exhibits category overlap and semantic ambiguity.
\rev{We adopt refinements already reported by \citet{Zeng2025First} and \citet{Li2024Understanding}, who analyse the CCS taxonomy itself and report evidence of type-level ambiguity from annotated CCS commits.}
\rev{Every type annotation is taken from the released \citet{Zeng2025First} corpus as published.}
\rev{Following these refinements, \texttt{perf} and \texttt{style} are absorbed into \texttt{refactor} and \texttt{chore} is excluded.}
This yields seven CCS types—\texttt{feat}, \texttt{fix}, \texttt{refactor}, \texttt{docs}, \texttt{test}, \texttt{build}, and \texttt{ci}—which serve as the label set used throughout dataset construction, model training, and evaluation.

\subsection{Tangled Commit Dataset}
\label{sc:met:dc}
There is no publicly available dataset of tangled commits annotated with CCS types.
Existing CCS datasets~\citep{Zeng2025First} contain only atomic commits, each representing a single CCS-style commit message that consists of a type, a description, and an associated code diff.

\rev{To work with tangled commits created from a reliable source of truly single-concern units, we used the manually annotated and verified CCS-labeled dataset released by \citet{Zeng2025First}.
We then constructed a pool of atomic commits, using Algorithm~\ref{alg:atomic_sampling}, considering the refined CCS types (Section~\ref{sc:met:ccs}), token lengths, and the repositories where each commit is originated from.}
This verified atomic pool serves as the compositional basis for synthesizing tangled commits used in both training and evaluation.

\begin{algorithm}
\SetKwInOut{Input}{Input}
\SetKwInOut{Output}{Output}

\Input{CCS-labeled Commit Dataset $\mathcal{R}$, \\
       Refined Label Set $\mathcal{T}$, \\
       Token-length Limit $l_\text{lim}$\rev{, \\
       Maximum Concern Count $n_{\max}$}}
\Output{Atomic Pool $\mathcal{A}$ grouped by \rev{repository and label}}

\rev{$\mathcal{R} \gets \textsc{MapToRefinedTypes}(\mathcal{R}, \mathcal{T})$}\\
$\mathcal{R} \gets \textsc{FilterByTokenLength}(\mathcal{R}, l_\text{lim})$\\
\rev{$P \gets \textsc{SelectRepositoriesByMinTypes}(\mathcal{R}, n_{\max})$}\\
\rev{$\mathcal{A} \gets \textsc{FilterByRepository}(\mathcal{R}, P)$}\\
\textbf{return} $\mathcal{A}$ \rev{grouped by repository and label}
\caption{Atomic Commit \rev{Pool Construction}}
\label{alg:atomic_sampling}
\end{algorithm}

\rev{Algorithm~\ref{alg:atomic_sampling} applies three deterministic filtering steps to the CCS-labeled dataset $\mathcal{R}$ in sequence.}
\rev{First, it maps each published type annotation onto the refined type set $\mathcal{T}$ (line 1).}
\rev{Second, it removes commits whose combined CCS description and code diff exceed the token-length limit $l_\text{lim}$ (line 2).}
\rev{Third, it selects repositories with at least $n_{\max}$ distinct types, where $n_{\max}$ is the maximum concern count in the tangled dataset (line 3), filters their commits into $\mathcal{A}$ (line 4), and returns $\mathcal{A}$ grouped by repository and label.}
\rev{We set $n_{\max} = 5$ because over 90\% of tangled commits involve three or fewer concerns (Section~\ref{sc:bg:tc}), and five covers the reported range and leaves room for the rarer, more heavily tangled cases.}
\rev{A repository whose commits cover only \texttt{fix}, \texttt{test}, and \texttt{docs} is therefore dropped, while one covering five or more of the refined types is kept.}
 
We then split the resulting atomic pool $\mathcal{A}$ into training and evaluation subsets using an \rev{80:20} ratio\rev{, keeping repositories disjoint across subsets,} to prevent data leakage.
Each subset serves as the input source for constructing the corresponding synthetic tangled dataset (either training or testing) in Algorithm~\ref{alg:tangled_generation}.

\begin{algorithm}
\SetKwInOut{Input}{Input}
\SetKwInOut{Output}{Output}

\Input{Atomic Pool $\mathcal{A}$ grouped by \rev{repository and} labels $\mathcal{T}$, \\
       Concern Counts $\mathcal{N}$, \\
       Per-count Quota $q$, \\
       Token-length Limit $l_\text{lim}$}
\Output{Tangled Commit Dataset $\mathcal{D}$}

$\mathcal{D} \gets \emptyset$\\
\ForEach{Concern Count $n \in \mathcal{N}$}{
    \While{Number of samples for count $n$ in $\mathcal{D}$ $<$ $q$}{
        Selected Label Set $T \gets \textsc{SelectDistinctLabels}(\mathcal{T}, n\rev{, \mathcal{D}})$\\
        \rev{Candidate Repositories $P_T \gets \{p : T \subseteq \textsc{DistinctTypes}(\mathcal{A}, p)\}$}\\
        \rev{Target Repository $r \gets \textsc{RandomSample}(P_T)$}\\
        Atomic Commits $C \gets \{\textsc{RandomSample}(\mathcal{A}\rev{_{r,t}}) : t \in T\}$\\
        Candidate commit $x \gets \textsc{ConstructTangledCommit}(C)$\\
        $x \gets \textsc{RemoveDuplicateCommits}(x, \mathcal{D})$\\
        $x \gets \textsc{FilterByTokenLength}(x, l_\text{lim})$\\
        \If{$x \neq \emptyset$}{
           $\mathcal{D} \gets \mathcal{D} \cup \{x\}$\\
        }
    }
}
\textbf{return} $\mathcal{D}$
\caption{Synthetic Tangled Commit Generation}
\label{alg:tangled_generation}
\end{algorithm}

In Algorithm~\ref{alg:tangled_generation}, 
each atomic commit in $\mathcal{A}$ provides a single CCS type, one commit message, and one code diff, allowing it to function as an independent compositional unit.
For a target concern count $n \in \{1,\dots,5\}$, \rev{\textsc{SelectDistinctLabels} selects $n$ distinct CCS types with the lowest frequencies in $\mathcal{D}$, with ties broken at random (line 4).}
\rev{Because all type frequencies are initially zero, the first label set is selected at random, after which less-used types are prioritized.}
\rev{$\textsc{DistinctTypes}(\mathcal{A}, p)$ is the set of types available from a repository $p$, so line 5 retains only a set of repositories $P_T$, each of which can supply all $n$ selected types.}
\rev{A target repository $r$ is then randomly drawn from $P_T$, and one atomic commit per selected type $t \in T$ is sampled from $r$ (lines 6-7).}
The algorithm constructs a tangled commit $x$ by merging their CCS types and concatenating their code diffs (line 8).
The resulting commit therefore contains a multi-label type set of size $n$, a combined CCS description, and an aggregated diff.
To maintain balanced coverage across concern levels, we set a per-count quota of $q = 350$, yielding $5 \times 350 = 1750$ tangled commits.
\rev{The tangled dataset comprises 1400 training and 350 evaluation commits.}
\rev{Label selection is constrained so that the seven types occur equally often in both splits.}
\rev{At every concern count the seven types contribute the same number of labels, from 50 each at $n=1$ to 250 each at $n=5$.}

Candidate samples exceeding the token-length limit or duplicating existing instances were discarded (i.e., represented as $\emptyset$), and sampling continued until all concern levels reached the target quota.

\begin{figure}
    \centering
    \includegraphics[width=0.75\linewidth]{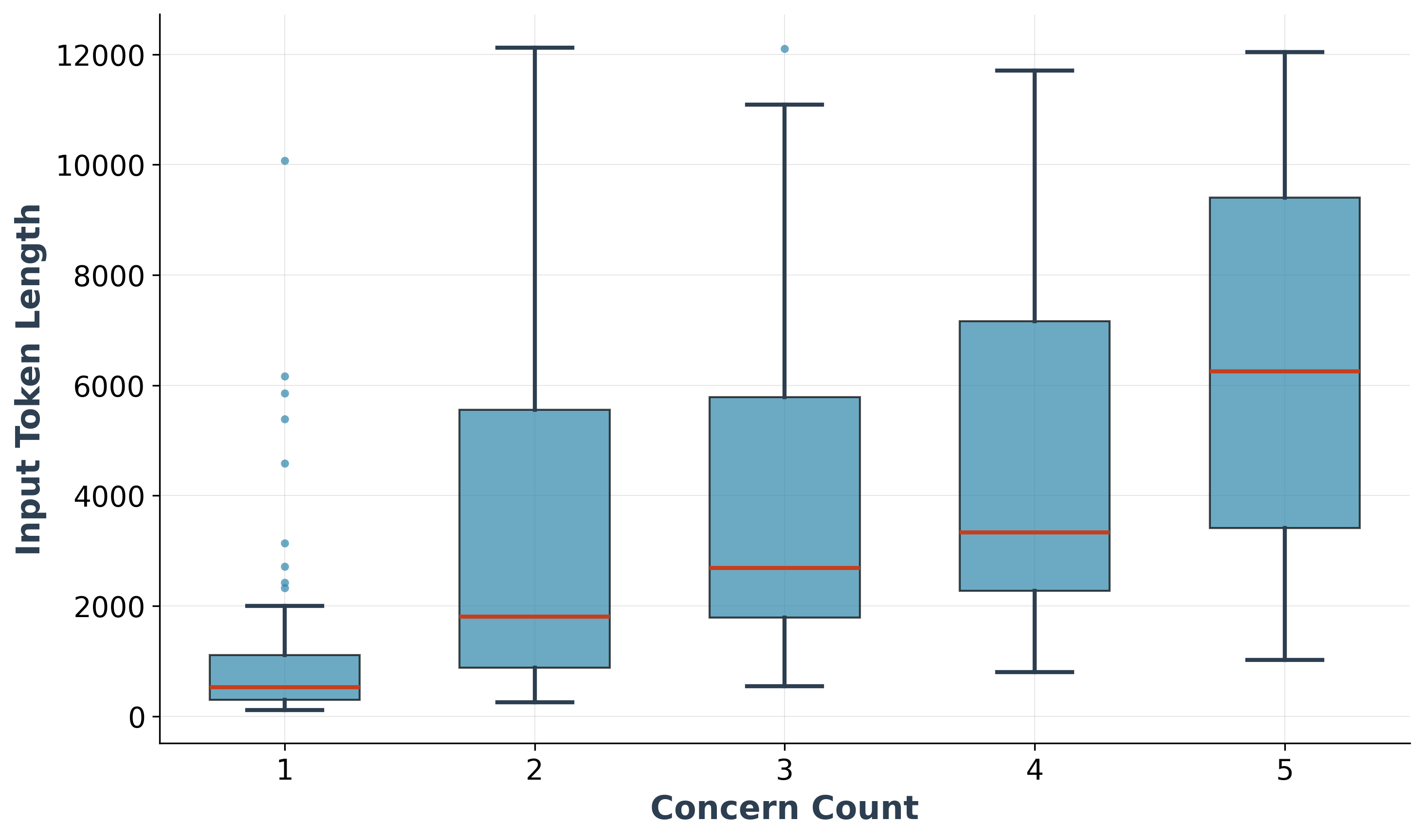}
    \caption{Token-length distribution across concern counts in the evaluation tangled dataset.}
    \label{fig:token_length_by_concern}
\end{figure}

Figure~\ref{fig:token_length_by_concern} presents the token-length distribution of the evaluation subset across concern counts.
Token length is measured as the combined number of tokens in the commit message and the associated code diffs.
As concern count increases, the distribution naturally shifts toward longer inputs because higher-count tangled commits aggregate multiple atomic changes in an approximately additive manner.

The resulting synthetic tangled dataset serves as the foundation for all subsequent stages of this study, providing the training subset used for model fine-tuning in Section~\ref{sc:met:model} and the evaluation subset used for performance assessment under the empirical setup described in Section~\ref{sc:evaluation}.

\subsection{Model Configuration}
\label{sc:met:model}

We select SLMs based on three task-aligned criteria: code awareness, context capacity, and reproducibility.
We restrict our choices to decoder-only Transformer architectures, as prior work shows that they provide strong performance and practical deployability for code-related tasks \citep{Lu2025Small}.
We further require that candidate SLMs be open and locally deployable, so that experiments remain reproducible and computationally efficient.
The context window must be large enough to contain the entire model input—both the prompt described in Section~\ref{sc:met:pd} and the tangled commit, whose token-length distribution is shown in Figure~\ref{fig:token_length_by_concern}.

We fine-tune the selected SLM using LoRA within the \rev{\texttt{Unsloth}} framework to specialize the model for our task.
The LoRA adapters are trained on the dataset introduced in Section~\ref{sc:met:dc}.
Inference is performed locally, with fixed random seeds to ensure reproducible outputs.

\subsection{Prompt Design}
\label{sc:met:pd}
\begin{figure}
    \centering
    \fbox{\includegraphics[width=0.75\linewidth]{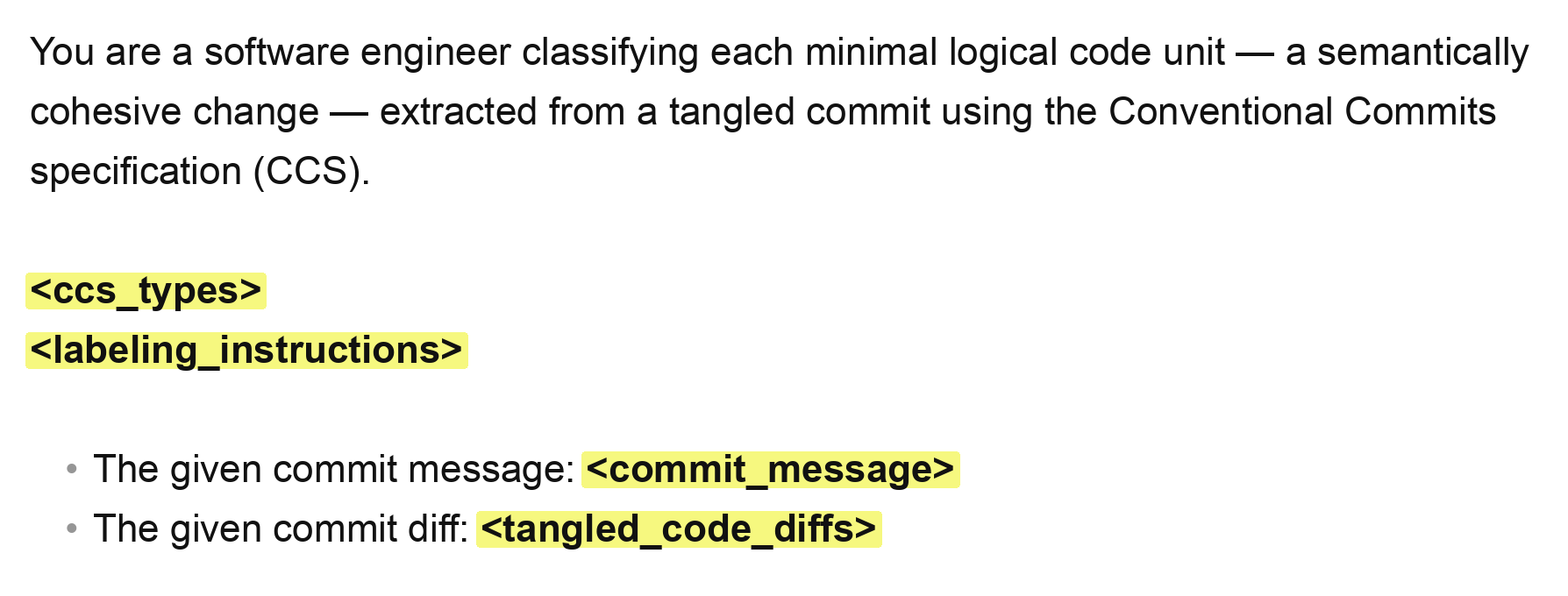}}
    \caption{Simplified structure of the prompt template used in the experiment.}
    \label{fig:prompt-template}
\end{figure}

We designed a structured prompt that defines the model’s role as a software engineer performing CCS concern classification.
It provides concise label definitions, deterministic decision rules, and minimal task context, as summarized in Figure~\ref{fig:prompt-template}.
The \texttt{<ccs\_types>} section adopts the refined types described in Section~\ref{sc:met:ccs}, ensuring consistent and unambiguous label usage across all experiments.
The \texttt{<labeling\_instructions>} section builds upon the \emph{purpose} and \emph{object} concepts introduced in Section~\ref{sc:bg:ccs}, extending the initial priority rule into a conditional scheme that explicitly resolves label overlap:

\begin{enumerate}[1.]
    \item \textbf{Purpose–Purpose:} select the label that best reflects \emph{why} the change was made.
    \item \textbf{Object–Object:} select the label that reflects the \emph{functional role} of the modified artefact.
    \item \textbf{Purpose–Object:} apply purpose labels only when the change affects application behaviour or structure.
\end{enumerate}

The prompt separates \emph{purpose} (\texttt{feat}, \texttt{fix}, \texttt{refactor}) from \emph{object} (\texttt{docs}, \texttt{test}, \texttt{build}, \texttt{ci}) and encodes these rules explicitly to ensure consistency in overlapping cases. 
Purpose labels are applied only when the modification alters behavior or structure; otherwise, the corresponding object label is used (e.g., editing comments in code is labeled as \texttt{docs}, not \texttt{refactor} or \texttt{fix}).
The input section comprises a \texttt{<commit\_message>} and \texttt{<tangled\_code\_diffs>}, both derived from the dataset described in Section~\ref{sc:met:dc}. 
Each tangled sample merges multiple atomic commit messages and their corresponding code diffs, which are subsequently varied during evaluation (Section~\ref{sec:exp:setup}) to examine the effects of message inclusion and token-budget-constrained diff truncation.
\section{Evaluation}
\label{sc:evaluation}

We aim to investigate whether SLMs can provide a practical and cost-effective alternative to LLMs for multi-label semantic concern detection by answering the four research questions discussed in Section~\ref{sec:intro}.

\begin{figure}[t]
    \centering
    \includegraphics[width=\linewidth]{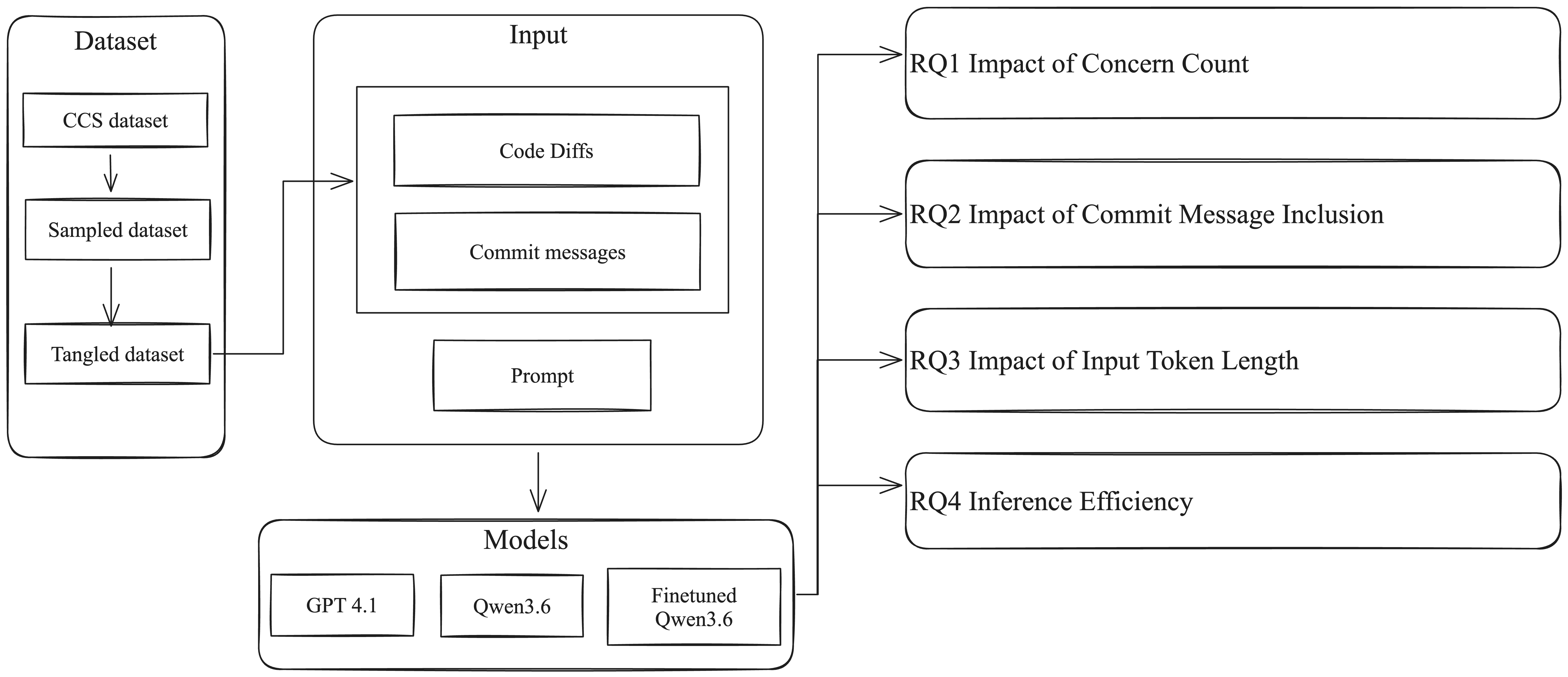}
    \caption{Evaluation Overview}
    \label{fig:rq-workflow}
\end{figure}

Figure~\ref{fig:rq-workflow} summarizes the empirical evaluation workflow, linking the datasets, model inputs, and research questions.
Based on the tangled commit dataset constructed in Section~\ref{sc:methodology}, we evaluate three configurations—GPT-4.1, Qwen3\rev{.6}-\rev{27B} (base), and Qwen3\rev{.6}-\rev{27B} (LoRA-fine-tuned)—on the multi-label concern detection task.
For each research question, we construct the corresponding model input by combining the prompt with either a full or truncated diff.
Commit messages are included by default in RQ1 and RQ3, compared between inclusion and exclusion in RQ2, and included again in RQ4 to quantify their marginal impact on IL (Qwen3\rev{.6}-\rev{27B} (LoRA)).
We then analyse the outputs with respect to concern count (RQ1), commit-message inclusion (RQ2), robustness under token-budget–constrained truncation (RQ3), and end-to-end inference efficiency (RQ4).

\subsection{Experimental Setup}
\label{sec:exp:setup}

We adopt Qwen3\rev{.6}-\rev{27B} as our primary SLM because it provides open weights, a context window large enough for our inputs, and competitive performance among publicly available models~\citep{Yang2025Qwen3}.
Its improved reasoning capability makes it suitable for the multi-step semantic decisions involved in tangled-commit classification.
We evaluate both the base model and a LoRA-fine-tuned variant (rank $r=32$, $\alpha=48$, dropout 0.05), which enables task-specific adaptation while maintaining deployability on high-end consumer GPUs. These versions are hereafter referred to as Qwen3\rev{.6} and Qwen3\rev{.6}-FT, respectively.
In addition, GPT-4.1 is used as \rev{a state-of-the-art LLM baseline} for multi-label semantic concern detection.

\rev{To the best of our knowledge, no prior approach formulates commit-level CCS detection as a multi-label classification task, and neither commit untangling (Section~\ref{sc:bg:tc}) nor single-label CCS classification~\citep{Zeng2025First,Li2024Understanding} yields a comparable output. We therefore compare Qwen3.6-FT against task-compatible baselines, the untuned Qwen3.6 and GPT-4.1, which receive the same input and produce the same multi-label output.}
Furthermore, we focus on decoder-only architectures to evaluate generative reasoning: models are prompted to generate chain-of-thought and label decisions in natural language rather than outputting fixed embeddings.
Encoder-only models (e.g., CodeBERT) are strong classifiers but are not designed for our prompt-based generative setting and therefore are outside the scope of this study.

We evaluate multi-label classification performance (RQ\numrange{1}{3}) using the Hamming loss (HL), defined as the fraction of misclassified labels per commit, averaged across all commits.
HL penalises both missing and spurious labels symmetrically, yielding an interpretable per-label error rate.
This property is well-suited to our setting: tangled commits activate varying subsets of the seven CCS types, and the number of active labels increases with the number of concerns.
As complexity increases, both over-prediction and under-prediction errors accumulate, making a stable, label-count-invariant metric desirable.
HL meets this requirement by applying a uniform penalty to each label error, independent of how many labels are active in a given commit.
In contrast, F1-based metrics are more sensitive to aggregation choices and less directly tied to per-label misclassification rates in this setting~\citep{Wu2020Multilabel}.
\rev{We report mean HL with 95\% confidence intervals~\citep{OBrien2016Interpret}.}

We measure efficiency (RQ4) using end-to-end inference latency (IL), defined as the wall-clock time from submitting the prompt to receiving the final output token~\citep{Chitty-Venkata2025LLMInferenceBench}.
This follows established industry practice\footnote{\url{https://docs.nvidia.com/nim/benchmarking/llm/latest/metrics.html}} and reflects user-visible deployment cost.
To account for non-determinism in generative inference, we run each configuration three times and remove outliers using the standard $1.5\times\mathrm{IQR}$ rule.
Frontier LLMs show run-to-run variability, and even the locally deployed SLM can exhibit occasional noise despite fixed seeds.

For RQ\numrange{1}{3}, we compare model configurations on the same set of commits using two-sided Wilcoxon signed-rank tests, and we report the resulting $p\text{-value}$s alongside the Vargha--Delaney effect size $\hat{A}_{12}$ to characterise both statistical and practical significance.
$\hat{A}_{12}$ quantifies the probability that the first model produces higher values than the second, with values below 0.5 indicating a tendency towards lower values~\citep{Arcuri2014Hitchhikers}.
For example, the values of $\hat{A}_{12} < 0.5$ indicate that the first model achieves lower HL than the second, meaning the first model is more accurate than the second.
For efficiency analysis (RQ4), we compute Pearson’s correlation coefficient ($r$).

All experiments are conducted on a \rev{single workstation equipped with an NVIDIA RTX PRO 6000 Blackwell GPU}.

\subsection{RQ1: Impact of Concern Count}
\label{sec:rq1}
\subsubsection{Setup}
RQ1 evaluates whether an SLM can approach a state-of-the-art LLM as the semantic complexity of a commit increases. 
To isolate the effect of concern count, we vary the number of semantic concerns combined to form each tangled commit ($n \in {1,2,3,4,5}$) while keeping all other conditions fixed.
Commit messages are always included, and the full diff content is supplied without truncation.
The single-concern case ($n=1$) serves as the atomic baseline.

\subsubsection{Results}
%
\begin{table}
  \centering
  \small
  \caption{\rev{Model performance in Hamming Loss across concern counts (mean with 95\% CI in brackets).}}
  \label{tab:cc-summary}

  \begin{tabular}{c rrr}
    \toprule
    \rev{$n$} & \rev{GPT-4.1} & \rev{Qwen3.6} & \rev{Qwen3.6-FT} \\
    \midrule
    \rev{1} & \rev{0.10 [0.07, 0.13]} & \rev{0.09 [0.06, 0.13]} & \rev{0.04 [0.02, 0.06]} \\
    \rev{2} & \rev{0.16 [0.12, 0.20]} & \rev{0.18 [0.13, 0.23]} & \rev{0.12 [0.08, 0.16]} \\
    \rev{3} & \rev{0.20 [0.16, 0.25]} & \rev{0.26 [0.20, 0.31]} & \rev{0.14 [0.10, 0.18]} \\
    \rev{4} & \rev{0.21 [0.17, 0.25]} & \rev{0.24 [0.20, 0.28]} & \rev{0.14 [0.10, 0.18]} \\
    \rev{5} & \rev{0.14 [0.11, 0.17]} & \rev{0.16 [0.13, 0.19]} & \rev{0.06 [0.03, 0.09]} \\
    \midrule
    \rev{All} & \rev{0.16 [0.15, 0.18]} & \rev{0.19 [0.17, 0.21]} & \rev{0.10 [0.09, 0.12]} \\
    \bottomrule
  \end{tabular}
\end{table}
As shown in \rev{Table~\ref{tab:cc-summary}}, \rev{Qwen3.6-FT} exhibits the lowest \rev{mean} HL \rev{(0.10), below GPT-4.1 (0.16) and base Qwen3.6 (0.19).}
\rev{Base Qwen3.6} shows \rev{a higher mean HL} \rev{than GPT-4.1,} whereas fine-tuning shifts the \rev{mean} \rev{below} GPT-4.1\rev{'s.}
This reduction in \rev{mean HL} relative to GPT-4.1 suggests that domain-specific fine-tuning \rev{allows} the \rev{SLM to surpass the frontier LLM on this task.}

\begin{table}
  \centering
  \small
  \caption{Pairwise comparisons across concern counts.
  $\hat{A}_{12} < 0.5$ indicates lower HL for the left-hand model, meaning the left-hand model is more accurate.}
  \label{tab:cc-pvalues}

  \begin{tabular}{c rr rr rr}
    \toprule
    & \multicolumn{2}{c}{GPT-4.1 vs Qwen3\rev{.6}}
    & \multicolumn{2}{c}{GPT-4.1 vs Qwen3\rev{.6}-FT}
    & \multicolumn{2}{c}{Qwen3\rev{.6} vs Qwen3\rev{.6}-FT} \\
    \cmidrule(lr){2-3} \cmidrule(lr){4-5} \cmidrule(lr){6-7}
    $n$ & $\hat{A}_{12}$ & $p\text{-value}$ & $\hat{A}_{12}$ & $p\text{-value}$ & $\hat{A}_{12}$ & $p\text{-value}$ \\
    \midrule
    1 & \rev{0.534} & \rev{0.258} & \rev{0.643} & \rev{$<$ 0.001} & \rev{0.609} & $<$ 0.001 \\
    2 & \rev{0.493} & \rev{0.650} & \rev{0.575} & \rev{0.019} & \rev{0.583} & \rev{0.034} \\
    3 & \rev{0.436} & $<$ 0.001 & \rev{0.605} & $<$ 0.001 & \rev{0.653} & $<$ 0.001 \\
    4 & \rev{0.440} & \rev{0.019} & \rev{0.617} & $<$ 0.001 & \rev{0.670} & $<$ 0.001 \\
    5 & \rev{0.445} & \rev{0.029} & \rev{0.668} & $<$ 0.001 & \rev{0.710} & $<$ 0.001 \\
    \bottomrule
  \end{tabular}
\end{table}

Table~\ref{tab:cc-pvalues} shows that \rev{Qwen3.6-FT is more accurate than both GPT-4.1 and base Qwen3.6 at every concern count} (all $p\text{-value}\le\rev{0.034}$).
\rev{By contrast, GPT-4.1 is not significantly more accurate than the base SLM at $n=1$ and $n=2$, and for $n\ge3$ it outperforms the base SLM with only small} effect sizes.
\rev{For $n \ge 2$, $\hat{A}_{12}$ against both baselines rises steadily with concern count, so a Qwen3.6-FT prediction is increasingly likely to have a lower HL than a baseline prediction as commits become more tangled.}
Fine-tuning yields \rev{consistent} improvements over the base SLM across all concern counts, confirming that task-specific adaptation provides substantial gains \rev{across the full range of} concern counts.


\rev{Table~\ref{tab:cc-summary}} reveals \rev{how HL shifts} across concern counts.
In the atomic case ($n=1$), Qwen3\rev{.6}-FT exhibits \rev{the lowest} central tendency of the \rev{three models,} and \rev{base Qwen3.6 (mean HL 0.09) very narrowly edges out GPT-4.1 (mean HL 0.10).}
\rev{For} $n\rev{\ge}2$ (the typical range reported in Section~\ref{sc:background}), the fine-tuned variant \rev{attains the lowest HL of the three models, preserving the ordering $\text{Qwen3.6-FT} \le \text{GPT-4.1} \le \text{Qwen3.6}$.}
Consistent with the \rev{lower mean HL}, the fine-tuned SLM shows substantially improved typical-case performance over the base model. When summarised using mean HL, this improvement corresponds to an error reduction on the order of \rev{\numrange{33}{62}\% at every concern count}---see the replication package (Section~\ref{sec:data-availability}) for detailed results.
Across models, \rev{mean HL rises from the atomic case towards a peak at $n=3$--$4$ and eases again at $n=5$, rather than degrading monotonically} with concern count.
The performance gap between fine-tuned and base models \rev{is largest around $n=3$--$4$, where task difficulty peaks} for \rev{all models.}

\begin{tcolorbox}
\textbf{Answer to RQ1:}
The answer to RQ1 is that multi-concern detection \rev{is hardest at three to four concerns and eases at five, rather than becoming progressively harder as concern count increases.}
Fine-tuning improves SLM reliability \rev{across all tested concern counts, with the fine-tuned SLM outperforming GPT-4.1 throughout this range.}
We therefore consider a fine-tuned SLM a practical default \rev{across this full range.}
\end{tcolorbox}

\subsection{RQ2: Impact of Commit Message Inclusion}
\label{sec:rq2}
\subsubsection{Setup}
RQ1 established how concern count affects model performance when full diffs and commit messages are available.
Building on that foundation, RQ2 investigates whether commit messages provide additional semantic cues beyond code diffs alone and, crucially, how well each model performs when messages are absent.
Real-world commit messages vary widely in quality, and many are incomplete, uninformative, or missing altogether.
To isolate this effect, we compare two settings: one in which the commit message accompanies the diff and one in which it is omitted.

All other aspects of the input remain fixed: full diff content is provided without truncation, and the analysis spans all concern-count levels to capture message effects across varying semantic complexity.

\subsubsection{Results}
%
\begin{table}
    \centering
    \small
    \caption{\rev{Impact of commit message inclusion on Hamming Loss (mean with 95\% CI in brackets).}}
    \label{tab:msgimpact-summary}

    \begin{tabular}{l rr}
      \toprule
      \rev{Model} & \rev{Without Msg} & \rev{With Msg} \\
      \midrule
      \rev{GPT-4.1} & \rev{0.18 [0.17, 0.20]} & \rev{0.16 [0.15, 0.18]} \\
      \rev{Qwen3.6} & \rev{0.21 [0.19, 0.22]} & \rev{0.19 [0.17, 0.21]} \\
      \rev{Qwen3.6-FT} & \rev{0.15 [0.13, 0.17]} & \rev{0.10 [0.09, 0.12]} \\
      \bottomrule
    \end{tabular}
\end{table}

\begin{table}
    \centering
    \small
    \caption{Impact of commit message on model performance.
    $\hat{A}_{12} > 0.5$ indicates a higher HL (i.e., less accurate) without the message.}
    \label{tab:msgimpact-pvalues}

    \begin{tabular}{p{3cm} rr}
      \toprule
      Model & $\hat{A}_{12}$ & $p\text{-value}$ \\
      \midrule
      GPT-4.1 & \rev{0.534} & \rev{$<$ 0.001} \\
      Qwen3\rev{.6} & \rev{0.532} & \rev{0.003} \\
      Qwen3\rev{.6}-FT & \rev{0.579} & $<$ 0.001 \\
      \bottomrule
    \end{tabular}
\end{table}

Table~\ref{tab:msgimpact-summary} shows the impact of commit message inclusion on detection accuracy across all models.
GPT-4.1 exhibits a small \rev{mean} HL reduction when messages are included \rev{(0.18 to 0.16). The base} Qwen3\rev{.6} shows a similar modest decrease \rev{(0.21 to 0.19),} while Qwen3\rev{.6}-FT shows a \rev{larger reduction (0.15 to 0.10, a 31\% relative} reduction).
As summarised in Table~\ref{tab:msgimpact-pvalues}, commit-message inclusion yields statistically significant improvements for all three models (all $p\text{-value}\le\rev{0.003}$).
Effect sizes ($\hat{A}_{12}$) reveal \rev{small} variation in magnitude: GPT-4.1 and the base SLM show only \rev{negligible} effects, whereas the fine-tuned SLM exhibits \rev{the largest} effect, indicating a \rev{stronger benefit from} explicit semantic cues provided by commit messages.

These results demonstrate that commit messages provide semantic cues that complement code diffs, with the effect being most pronounced for fine-tuned SLMs.
For Qwen3\rev{.6}-FT, messages provide explicit semantic cues that help the model disambiguate multi-label classification tasks.
By contrast, GPT-4.1's stable performance suggests it can effectively infer intent from diffs alone, though messages still provide measurable benefit.
Commit messages, therefore, represent a critical input modality, particularly for resource-efficient SLMs.

While message inclusion consistently reduces HL across models, the improvement is most pronounced for SLMs, especially Qwen3\rev{.6}-FT, where error rates drop \rev{most}.
Practically, messages should be enabled by default in deployments, with \rev{the largest} HL gains expected for resource-efficient SLMs.

\begin{tcolorbox}
\textbf{Answer to RQ2.}
Commit-message inclusion consistently reduces HL across models, with the largest gains observed for the fine-tuned SLM. 
GPT-4.1 can often infer intent from diffs alone, but still benefits from messages, whereas the SLM relies more strongly on these explicit semantic cues.
\end{tcolorbox}

\subsection{RQ3: Robustness Under Token-Budget–Constrained Diff Truncation}
\label{sec:rq3}
RQ1 showed that concern count drives task difficulty, and RQ2 demonstrated that commit messages provide useful semantic cues.
RQ3 examines a different practical constraint: how reliably an SLM can detect multiple concerns when the token budget is reduced, and different segments must be partially truncated. Such conditions are common in deployments of SLMs, where limited context windows and multi-concern commits compete for input space.

To evaluate robustness under constrained input budgets, we vary the maximum number of tokens that can be allocated to the commit ($L \in {1024, 2048, 4096, 8192, 12288}$).
Commit messages are always included, and all concern-count levels are evaluated.

A header-preserving truncation policy is applied: diff headers and the earliest modified hunks are retained first, and trailing content is removed only when the budget is exceeded.
Let $|p|$ and $|m|$ denote the token lengths of the preserved diff prefix and the commit message, respectively. 
We define the remaining token budget for diff segments as \begin{equation} L' = \max(L - |m| - |p|, 0). \end{equation}
The remaining capacity is divided evenly across the $n$ diff segments, and any segment exceeding its allocated share is truncated from the tail.
This setup allows us to isolate the effect of selective, segment-level truncation without exceeding the model’s global context window.

\subsubsection{Results}
%
\begin{table}
  \centering
  \small
  \caption{\rev{Impact of token budget on Hamming Loss for different models (mean with 95\% CI in brackets).}}
  \label{tab:ctxlen-summary}

  \begin{tabular}{c rrr}
    \toprule
    \rev{$L$} & \rev{GPT-4.1} & \rev{Qwen3.6} & \rev{Qwen3.6-FT} \\
    \midrule
    \rev{1024} & \rev{0.15 [0.14, 0.17]} & \rev{0.17 [0.15, 0.19]} & \rev{0.12 [0.10, 0.14]} \\
    \rev{2048} & \rev{0.16 [0.14, 0.17]} & \rev{0.18 [0.16, 0.19]} & \rev{0.11 [0.10, 0.13]} \\
    \rev{4096} & \rev{0.16 [0.14, 0.17]} & \rev{0.19 [0.17, 0.21]} & \rev{0.11 [0.09, 0.13]} \\
    \rev{8192} & \rev{0.16 [0.14, 0.17]} & \rev{0.19 [0.17, 0.21]} & \rev{0.10 [0.08, 0.12]} \\
    \rev{12288} & \rev{0.16 [0.15, 0.18]} & \rev{0.19 [0.17, 0.21]} & \rev{0.10 [0.09, 0.12]} \\
    \bottomrule
  \end{tabular}
\end{table}

\begin{table}
  \centering
  \small
  \caption{Pairwise comparisons across token budgets.
  $\hat{A}_{12} < 0.5$ indicates lower HL for the left-hand model, meaning the left-hand model is more accurate.}
  \label{tab:ctxlen-pvalues}

  \begin{tabular}{c rr rr rr}
    \toprule
    & \multicolumn{2}{c}{GPT-4.1 vs Qwen3\rev{.6}}
    & \multicolumn{2}{c}{GPT-4.1 vs Qwen3\rev{.6}-FT}
    & \multicolumn{2}{c}{Qwen3\rev{.6} vs Qwen3\rev{.6}-FT} \\
    \cmidrule(lr){2-3} \cmidrule(lr){4-5} \cmidrule(lr){6-7}
    $L$ & $\hat{A}_{12}$ & $p\text{-value}$ & $\hat{A}_{12}$ & $p\text{-value}$ & $\hat{A}_{12}$ & $p\text{-value}$ \\
    \midrule
    1024 & \rev{0.474} & $<$ 0.001 & \rev{0.564} & $<$ 0.001 & \rev{0.588} & $<$ 0.001 \\
    2048 & \rev{0.481} & \rev{0.015} & \rev{0.588} & $<$ 0.001 & \rev{0.606} & $<$ 0.001 \\
    4096 & \rev{0.464} & $<$ 0.001 & \rev{0.594} & $<$ 0.001 & \rev{0.625} & $<$ 0.001 \\
    8192 & \rev{0.462} & $<$ 0.001 & \rev{0.608} & $<$ 0.001 & \rev{0.640} & $<$ 0.001 \\
    12288 & \rev{0.472} & $<$ 0.001 & \rev{0.618} & $<$ 0.001 & \rev{0.638} & $<$ 0.001 \\
    \bottomrule
  \end{tabular}
\end{table}

Table~\ref{tab:ctxlen-summary} reveals that model performance remains largely stable across token lengths.
GPT-4.1 shows essentially stable performance across all budgets, with \rev{mean} HL remaining \rev{between 0.15} and \rev{0.16.}
\rev{Base Qwen3.6} exhibits similar stability, with HL fluctuating between \rev{0.17} and \rev{0.19} across all lengths.
Qwen3\rev{.6}-FT shows \rev{the lowest} HL \rev{at every budget,} ranging from \rev{0.12 at the smallest budget ($L=1024$) down} to \rev{0.10 at the two largest budgets ($L=8192$ and $L=12288$), a slight improvement} as \rev{the token budget grows.}
As summarised in Table~\ref{tab:ctxlen-pvalues}, all pairwise comparisons remain statistically significant across token budgets (all $p\text{-value}\rev{\le0.015}$).
Effect sizes ($\hat{A}_{12}$) remain highly stable across token budgets: GPT-4.1 \rev{outperforms the base SLM with negligible effects, the fine-tuned SLM outperforms GPT-4.1 with small effects, and fine-tuning consistently outperforms the base SLM with small-to-medium effects.}
This consistency indicates that token-budget–constrained truncation does not materially alter the relative ordering between model configurations.

These results indicate that detection accuracy is highly robust to partial truncation of diff segments under our header-preserving policy. 
Even under the smallest budget ($L{=}1024$), where truncation is most pronounced, all models perform comparably to the full-budget configuration ($L{=}12288$).
In other words, truncating diff segments from the tail while retaining headers and early modified hunks does not materially degrade multi-concern detection performance, even when a substantial fraction of the diff tokens is removed.

\begin{tcolorbox}
\textbf{Answer to RQ3:}
The answer to RQ3 is that, under a header-preserving truncation policy, detection accuracy remains stable across token budgets from \numrange{1024}{12288} tokens.
In our setting, commit messages and early diff regions provide sufficient semantic signal for SLMs to maintain multi-concern detection performance without full diff coverage, so deployments can reduce token budgets without substantial loss in accuracy.
\end{tcolorbox}

\subsection{RQ4: Inference Efficiency}
\subsubsection{Setup}
In RQ4, we focus on the Qwen3\rev{.6}-FT model and quantify how concern count, commit-message inclusion, and token budget affect inference latency (IL).
\rev{We report IL for the locally deployed model only. Cloud API latency depends on network and provider-side conditions, whereas local inference depends on the local hardware configuration. A direct comparison would therefore be deployment-specific.}
To ensure consistency with earlier research questions, we use the same dataset, experimental configurations, and token budget constraints as defined in Sections \ref{sec:rq1}–\ref{sec:rq3}.


\subsubsection{Results}

\paragraph{Experiment 1: Impact of Concern Count}
\begin{figure}[t]
\centering
\includegraphics[width=\linewidth]{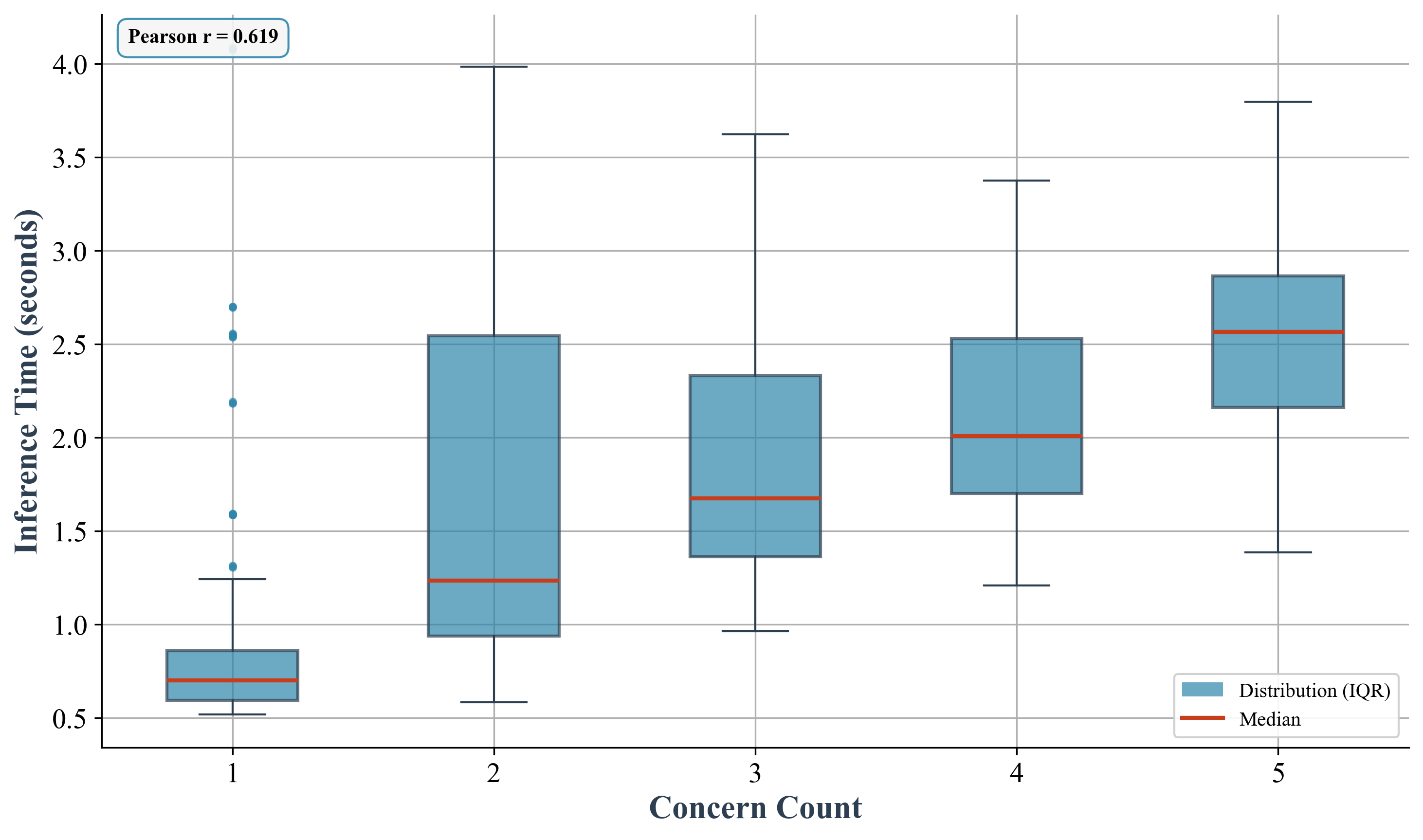}
\caption{Inference time by concern count}
\label{fig:regression_concern_count}
\end{figure}

Figure~\ref{fig:regression_concern_count} shows that median IL rises monotonically as concern count increases from one to five. Pearson correlation is strong \rev{($r=0.619$)}, consistent with a monotonic increase in IL as concern count grows.
The interquartile range widens at higher concern levels, indicating increased computational load.
These results establish concern \rev{count} as a primary driver of inference latency.

\paragraph{Experiment 2: Impact of Message Inclusion}
\begin{figure}
\centering
\includegraphics[width=0.9\linewidth]{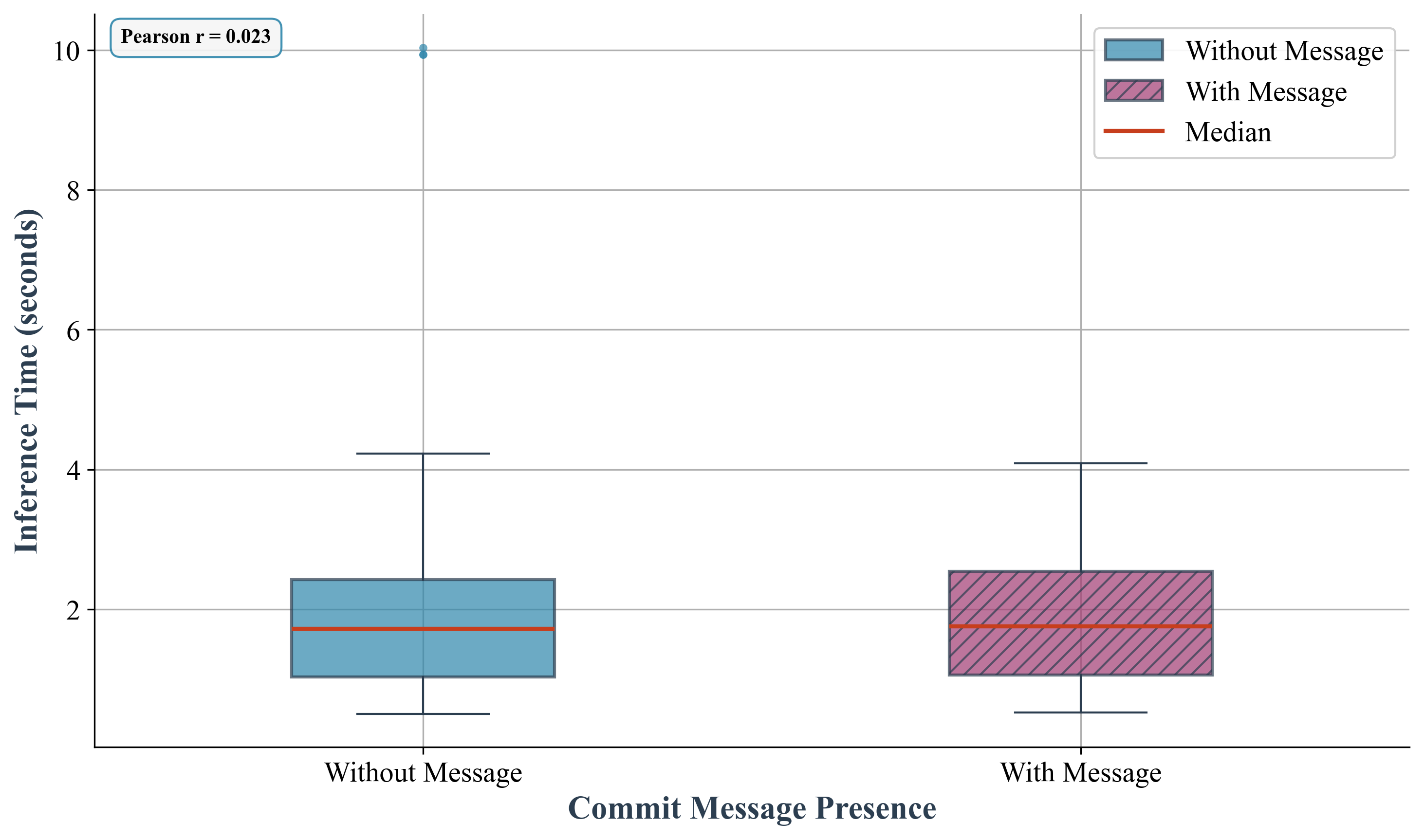}
\caption{Inference time by commit message inclusion}
\label{fig:commit_message_time_comparison}  
\end{figure}

Figure~\ref{fig:commit_message_time_comparison} shows overlapping box plots with nearly unchanged medians. Including commit messages added only \rev{${\sim}0.04$s} of latency, a negligible overhead.
This weak association aligns with transformer execution patterns: message tokens form a short auxiliary segment appended to diffs, contributing marginally to the total input length.
This small increase is consistent with the fact that commit messages add relatively few tokens compared with diffs, and therefore contribute only marginally to overall input length and observed IL in our setting.
From a deployment perspective, messages should be enabled by default for their semantic benefits (as shown in RQ2) rather than disabled for efficiency, as their IL impact is negligible.
\paragraph{Experiment 3: Impact of Token Budget}
\begin{figure}[t]
\centering
\includegraphics[width=0.9\linewidth]{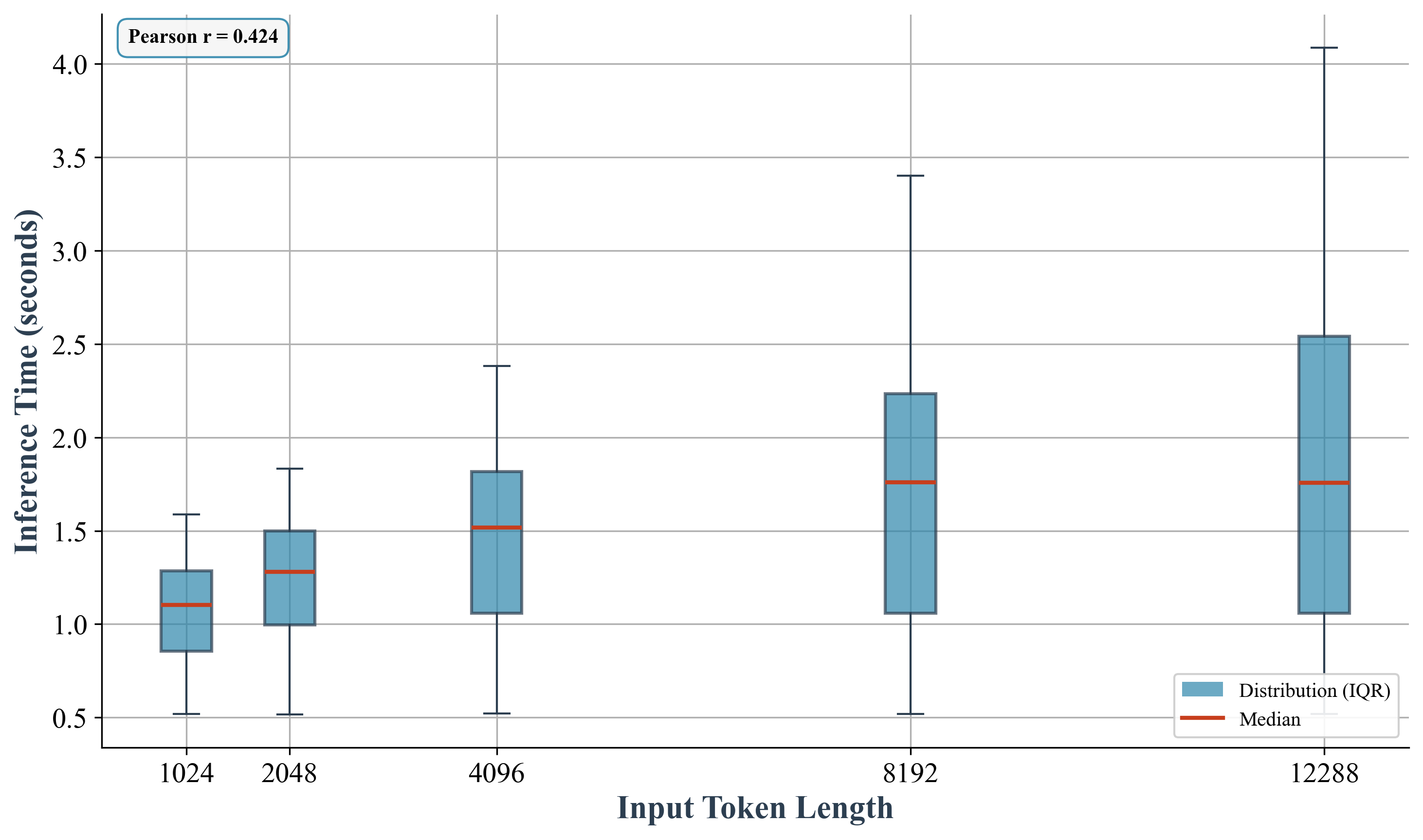}
\caption{Inference time by token budget}
\label{fig:input_tokens_time_comparison}
\end{figure}

Figure~\ref{fig:input_tokens_time_comparison} shows \rev{IL increasing with input token length} across \rev{the tested budgets.}
Correlation analysis reports a \rev{moderate} positive association (Pearson \rev{$r=0.424$}). 
While the computational cost of attention grows with input length~\citep{Vaswani2017Attention}, the observed effect on IL is modest relative to concern count.
This suggests that token-length management (through truncation or summarization) provides incremental rather than substantial efficiency gains, making it a complementary but secondary optimization strategy.

\begin{tcolorbox}
\textbf{Answer to RQ4:}
Inference latency scales with the total number of input tokens, which naturally increases as concern count grows.
Commit messages provide a cost-effective accuracy boost with negligible latency overhead, whereas diff size (which tends to grow with concern multiplicity) is the dominant practical driver of IL in our experiments.
\end{tcolorbox}

\subsection{Threats to Validity}
\label{sec:eval:threats}
We reduce construct validity risks by limiting the CCS taxonomy per prior refinements~\citep{Li2024Understanding,Zeng2025First}\rev{, as described in Section~\ref{sc:met:ccs}}.
\rev{The distinctions we set aside are the ones on which annotators agree least, and that ambiguity is plausibly a cause of tangling rather than an incidental property of the taxonomy. The retained types are not equally demanding either. \texttt{test}, \texttt{docs}, \texttt{build}, and \texttt{ci} are often identifiable from file paths alone, whereas \texttt{feat}, \texttt{fix}, and \texttt{refactor} depend on the intent behind the edit, so aggregate scores may overstate how far the models recover semantic concerns. Our results therefore do not estimate performance over the full CCS taxonomy.}
At the same time, the synthetic tangled dataset yields a controlled benchmark in which concern count and token budget can be varied systematically, allowing us to study their effects under well-defined conditions.

\rev{Hamming loss averages errors uniformly over the seven labels, so a model that never predicts an infrequent type can still record a low loss. This failure mode cannot arise in our evaluation because the generation procedure (Section~\ref{sc:met:dc}) makes every label occur equally often in both splits, so no type is infrequent. That uniformity is a property of the benchmark rather than of real repositories, and evaluating the same models on naturally distributed commits remains future work.}

A key threat arises from stochastic variation in generative inference. We therefore repeat each experimental configuration three times and aggregate across runs, which mitigates the effect of sporadic runtime noise on both the frontier LLM and the local SLM.

External validity is limited by our use of a synthesized tangled dataset rather than naturally tangled industrial commits. \rev{Because each tangled instance is produced by concatenating the diffs of atomic commits, the resulting tangles do not reproduce the tighter entanglement observed in practice: concerns modifying the same statements, sharing variables, or interacting through call-graph dependencies. The constituent atomic commits may also originate from different points in a project's history, so we cannot guarantee that they would have been authored as part of a single development task.} \rev{Following prior studies of tangled code changes~\citep{Herzig2011Untangling, Dias2015Untangling}, we additionally examined whether any two constituent concerns of a tangled commit modify the same file or the same directory. Only a small fraction of tangled commits show file-level overlap, while directory-level overlap is more frequent. The full measurements are included in our replication package (Section~\ref{sec:data-availability}). Since no reference distribution exists and not all tangled commits are necessarily interleaved in file or directory, these observations provide a descriptive account of the dataset's co-location properties.} To mitigate this risk, the atomic pool is drawn from a verified CCS corpus~\citep{Zeng2025First}, and samples are stratified by concern count with uniform label sampling, ensuring balanced coverage of concern multiplicity even when natural label frequencies vary across repositories. The atomic commits span multiple projects and languages. \rev{Evaluation} on naturally tangled commits would further strengthen external validity.

\rev{Our detector operates at the granularity of the whole commit: it reports which semantic concerns a commit contains, but not which changed lines or hunks belong to each concern. Its output therefore cannot by itself drive tasks that require line-level attribution, such as automatically splitting a tangled commit into single-concern commits or routing individual hunks to the reviewers responsible for each concern. Recovering that mapping requires a partitioning step, which we do not attempt here; whether concern labels predicted at the commit level can supervise or constrain such a step is left to future work.}

A further external validity threat concerns RQ3.
As shown in Figure~\ref{fig:token_length_by_concern}, the token-length distribution of our CCS-derived dataset spans a wide range, meaning that under our token-budget settings some commits are truncated while others remain within budget.
Our robustness results therefore reflect this mixture and are specific to the header-preserving truncation policy we adopt, which always retains file headers and early modified hunks.
We do not examine alternative truncation schemes that remove these regions or truncate from the head or middle of the diff, and repositories with systematically longer or more fragmented diffs may behave differently under the same budgets.
Evaluating such alternative policies and longer-horizon histories is an important direction for future work.

\subsection{Data Availability}\label{sec:data-availability}
All artefacts produced in this study, including the tangled commit dataset, \rev{a} fine-tuned \rev{LoRA adapter}, and supporting scripts, have been made publicly available to ensure reproducibility.
The tangled dataset used for evaluation is hosted on Hugging Face\footnote{\url{https://huggingface.co/datasets/Berom0227/tangled-ccs-commits}}.
\rev{The} fine-tuned \rev{model is} released on Hugging Face \rev{as a LoRA adapter}\footnote{\href{https://huggingface.co/Berom0227/Semantic-Concern-SLM-Qwen3.6-27B-adapter}{\nolinkurl{https://huggingface.co/Berom0227/Semantic-Concern-SLM-Qwen}\rev{\nolinkurl{3.6-27B}}\nolinkurl{-adapter}}}.
Supporting scripts, training configurations, evaluation pipelines, and extended result tables are archived on GitHub\footnote{\url{https://github.com/GoBeromsu/Detecting-Multiple-Semantic-Concerns-in-Tangled-Code-Commits-using-Small-Language-Models}},
providing complete resources for reproducing the experimental results. 

\section{Discussion}
\label{sc:discussion}

\subsection{Detecting Multiple Concerns using SLM}
\label{sc:ds:slm}
Tangled commits arise at the moment a commit is created, which places detection directly after the commit step within the VCS workflow. 
Because committing is a daily and iterative activity for developers, any process coupled to this step inherits the same practical constraints.
This includes the fact that a codebase constitutes a core organisational asset whose security and privacy requirements prevent repository data from being processed outside the VCS environment. 
Consequently, tangled commit detection must be cost-sensitive, latency-sensitive, readily automatable, and executable entirely within local infrastructure to function effectively in real development contexts.
As discussed in Section \ref{sc:evaluation}, the detection task itself remains structurally constrained and simple: the model receives a tangled commit and is required to classify a predefined multi-label output.

Given these constraints, the detection component must operate as an internal module within the VCS workflow, and small language models align particularly well with this requirement. 
The CCS provides a machine-readable structure that enables SLMs to interpret semantic concerns consistently, reducing dependence on complex reasoning or large context windows. 
Their lightweight nature allows deployment directly within local or on-premise infrastructure, naturally satisfying the security and privacy constraints that prevent code from leaving private repositories. 
Moreover, SLMs maintain predictable latency and can be invoked efficiently within CI/CD pipelines, making them a practical fit for commit-level automation where stability, responsiveness, and cost-efficiency are essential.

\subsection{LLM vs SLM: What to Use in Practice}
\label{sc:ds:llmvslm}
Tangled commit detection varies in difficulty depending on the structure of the input.
Our experiments show three consistent patterns.
First, the problem becomes harder as the number of concerns increases.
Second, commit messages reliably improve performance, although the model can still reason without them.
Third, within our CCS-derived dataset, reducing the available token budget has little effect on difficulty, because most commits are short and our header-preserving truncation policy retains the most informative parts of each diff.

For model selection, although LLMs provide broader context windows and stronger reasoning capabilities, prior work \cite{Belcak2025Small} shows that such capacity is often excessive for a repetitive, structurally constrained task like tangled commit detection.
Our results reinforce this view. \rev{The fine-tuned SLM achieves the lowest HL at every concern count and token budget.}
Inference latency scales primarily with the total number of input tokens—which naturally grows with concern count—while commit messages offer a cost-effective accuracy boost with negligible latency overhead.
With appropriate input design, SLM-based detection can be integrated directly into VCS pipelines, and tangled commits can be handled efficiently by \rev{fine-tuned} SLMs.

\subsection{Open Challenges}
There are also challenges that remain unresolved for SLM-based multi-concern detection. 
Although LLMs offer larger context windows and stronger general-purpose reasoning, prior work \cite{Belcak2025Small} shows that tasks built around fixed input formats and repetitive decision patterns gain little from LLM-scale capacity. Our multi-concern commit classification task follows this structure in practice, with short diffs, stable input layouts, and bounded label granularity. Under these conditions, SLMs are not only sufficient but often the most appropriate choice.

\rev{Semantic concern classification and syntactic untangling are complementary rather than competing. Untangling partitions a change into structurally cohesive groups without naming the intent behind each group, whereas our task names the concerns present without localizing them. Bridging them requires extending semantic labelling to the code level, so that each changed fragment carries a concern type. A model performing this \emph{concern attribution} (Section~\ref{sc:bg:tc}) subsumes both tasks. Whether to label before partitioning, partition before labelling, or infer both jointly, and how to evaluate the result when neither component is exact, remain open.}

On the modelling side, it is not yet clear what constitutes an optimal SLM configuration, as alternatives such as quantised large models, varied prompting strategies, retrieval-augmented contexts, or different fine-tuning schemes may yield substantially different behaviours.
Furthermore, the question of how to construct or curate training datasets remains unsettled, as it is not yet clear which commit characteristics, tangling patterns, or sampling strategies are most appropriate for fine-tuning models on this task.
As discussed in Section~\ref{sc:ds:slm}, integrating the tangled commit detection into the VCS workflow raises several open design questions, including how such a component should be orchestrated, how fallback mechanisms ought to be structured, and how reliably the system would operate once deployed.
Extending this integration towards automated untangling also remains unresolved, particularly for commits with unclear or overlapping semantic boundaries.
\section{Related work}
\label{sc:related-work}
\subsection{Commit-Related Tasks}

Commit-level research is commonly grouped into three areas: \emph{generation}, \emph{classification}, and \emph{refinement}.
Generation tasks include commit message generation (CMG), code change summarization, code review generation (CRG), and just-in-time comment updates (JITCU). 
Classification typically targets the recovery of semantic intent, often via CCS types.
Refinement leverages commit context for code improvement and defect prediction.
Comparative studies report that LLMs achieve strong performance across CMG, CRG, and JITCU under in-context learning (ICL) and PEFT, albeit with substantial computational and latency costs \citep{Fan2024Exploring}.
Fine-tuned SLMs trained on high-quality data demonstrate competitive performance and offer a cost-efficient alternative \citep{Li2024Understanding}.
Code-change-oriented pretraining further improves generation quality.
For example, CCT5, pre-trained on \emph{CodeChangeNet} (1.5M commit–message pairs), consistently outperforms earlier approaches \citep{Lin2023CCT5}.
Refinement tasks extend these gains by exploiting commit context for automated refactoring and just-in-time defect prediction \citep{Li2024Understanding, Lin2023CCT5}.
However, most CCS-based classification assumes a single dominant intent per commit, leaving multi-intent (tangled) commits underexplored and motivating our multi-label formulation.
\rev{An exception is \citet{Sarwar2020Multilabel}, who allow multiple maintenance-activity labels per commit but predict them from the commit message alone. We target CCS-aligned concerns from both the message and the code diff in tangled commits.}

\subsection{Optimizing SLMs for Code-level Tasks}
Although our methodology achieves strong baseline performance using a curated dataset and CoT prompting, its capabilities can be further enhanced by incorporating techniques from recent SLM research.
To mitigate the limitation that language models often identify what changed but fail to infer why, Retrieval-Augmented Generation (RAG) can supply explicit, relevant examples during inference \citep{Li2024Understanding}.
More sophisticated prompt-engineering strategies also offer promising directions for improving reasoning on complex code changes \citep{Hou2024Large}.
Recent studies further suggest that SLM reasoning ability can be enhanced through distillation.
Although CoT prompting elicits reasoning in LLMs, such capabilities typically emerge only in very large models.
\citet{Ranaldi2024Aligning} show that instruction-tuned student models trained on a small number of high-quality distilled CoT demonstrations can recover much of this reasoning ability at smaller scales.
In particular, CoT-based demonstrations are most effective for reasoning alignment, especially when derived from in-family teacher models.
These findings indicate that SLMs, despite their parameter constraints, can acquire multi-step reasoning skills through carefully curated distillation rather than prompting alone.
\section{Conclusion}
\label{sc:conclusion}
To the best of our knowledge, this paper presents the first empirical study of multi-label semantic concern detection in tangled commits using SLMs.
We refine the CCS taxonomy, construct a controlled-synthetic dataset of tangled commits, and evaluate a fine-tuned \rev{27B}-parameter SLM against GPT-4.1 while varying concern count, commit-message inclusion, and token–budget–constrained diff truncation.
Our results show that multi-concern detection is feasible with SLMs at practical latency and cost, with \rev{detection} error \rev{generally} increasing \rev{with concern count;} the fine-tuned SLM \rev{outperforms} GPT-4.1 \rev{at every tested concern count}.
Commit messages yield consistent HL improvements with negligible latency cost, whereas token-budget–constrained inputs exhibit stable performance under header-preserving truncation, indicating that early diff context contains sufficient semantic signal.
Latency is primarily driven by the number of concerns, with minimal overhead from commit messages or token-budget tuning.
Preserving diff headers and early hunks appears more beneficial than enlarging context windows, indicating that effort is better spent on simple, robust preprocessing than on scaling context length.
In addition, commit messages should be enabled by default, as they provide substantial accuracy gains with negligible latency cost.
Domain-specialised SLMs are \rev{a} reliable \rev{default across the tested concern counts.}

Future work will examine how individual concern types differ in detection difficulty, how semantic cues interact as the number of concerns increases, and which specific diff or message features most strongly influence detection outcomes. 
We also plan to extend the evaluation to naturally tangled industrial histories and explore models that propose untangled commit partitions, moving beyond detection toward the longer-term goal of assisting developers in reconstructing tangled commits into more atomic units.

\section*{Declarations}

\subsection*{Funding}
The authors did not receive support from any organization for the submitted work.

\subsection*{Ethical Approval}
Not applicable.

\subsection*{Informed Consent}
Not applicable.

\subsection*{Author Contributions}
Beomsu Koh: Conceptualization, Methodology, Software, Formal analysis, Writing -- original draft.\\
Neil Walkinshaw: Supervision, Writing -- review \& editing.\\
Donghwan Shin: Supervision, Writing -- review \& editing.

\subsection*{Data Availability Statement}
All artefacts produced in this study, including the tangled commit dataset, \rev{a} fine-tuned \rev{LoRA adapter}, and supporting scripts, have been made publicly available. See Section~\ref{sec:data-availability} for details.

\subsection*{Conflict of Interest}
The authors have no conflicts of interest to declare that are relevant to the content of this article.

\subsection*{Clinical Trial Number}
Not applicable.

\bibliographystyle{spbasic}      
\bibliography{references} 


\end{document}